# Systematization of Equilibrium and Unstable Bloch Domain Walls in Plates of Cubic Ferromagnetic Crystals

**B.M. Tanygin[a1], O.V. Tychko[a]**

[a] Kyiv Taras Shevchenko National University, Radiophysics Faculty, Glushkov av.2, build.5, Kyiv, Ukraine, MSP 01601

[1]*Corresponding author:* B.M. Tanygin, 64 Vladimirskaya str., Taras Shevchenko Kyiv National University, Radiophysics Faculty. MSP 01601, Kyiv, Ukraine.

*E-mail*: b.m.tanygin@gmail.com

*Phone*: +380-68-394-05-52

**Abstract.**

The energy and structure of planar Bloch domain walls with opposite directions (right- and left-handed) and different paths of magnetization vector rotation have been systemized. The theory covers arbitrary oriented plates of cubic $m\bar{3}m$ symmetry magnetically ordered medium with negative anisotropy. The general approach is created and applied to most plates typically used in the experiments.

## 1. Introduction

The first important path of the process of the inhomogeneous magnetization reversal (IMR) in the real crystals is the domain nucleation from the local magnetization inhomogeneity (0° domain wall). These processes have been investigated in details [1]. The appearing of the new phase domain nucleus in the volume of the domain walls (DWs) is another important mechanism of IMR processes [2]. Then, the growth of the DW width $\delta$ [3] is the important mechanism of the magnetic reversal. It was determined that specific Bloch DW orientations correspond to the “infinity” width of the DW [3]. Consequently, general investigation of the DW structure as well as all possibilities of its changes (including nucleation and DW decay in the limit case) is the important problem for fundamental research and applications of the magnetically ordered media.

There are many works on the investigations of the DW structure [3-11]. Planar Bloch DWs [4] have been investigated in the unrestricted crystals [3-6] as well as in the restricted bulk (i.e., in cases when the crystal sample width is sufficiently larger [5] than DW one) crystals [6-11] with the negative cubic magnetocrystalline anisotropy. The spatial restriction of the volume of magnetically ordered medium leads to the reflection of the DW plain from its orientation which were realized in the unrestricted crystal and to growth of their energy density in general case [6,9,11]. In fact, only very specific cases of DWs were investigated in details for cases of bulk plates. They are mainly DWs with lowest energies in high-symmetrical oriented plates: (110) [6,9-11], (100) [12-14] and (111) [15,16]. The sequel analysis of domain structure of these samples requires investigation of all possible DWs in their volume. The same investigation are also required for sample with remained orientations (examples are widely used (112) [17] and (210)- [18] plates and films) as well as disoriented samples [6].

In our opinion, general and detailed systematization of the DWs in the restricted samples (plates) was not provided yet. General symmetry classification of the magnetic DWs was done [19,20]. To complete these investigations it is necessary to solve variation problem to found out numerically the thermodynamically equilibrium state of the DW structure for each case of the DW symmetry.

The aim of this work is building of the systematization of energy and structure of planar Bloch DWs with opposite directions (right- and left-handed) and different paths of magnetization vector rotation. The chosen constructive example of this approach is the consideration of the arbitrary oriented plates of cubic m$\bar{3}$m ferromagnetic crystals with negative magnetocrystalline anisotropy.

## 2. Directions and paths of magnetization rotation

Let $\mathbf{m}$ is ort along $\mathbf{M}$ in the DW volume: $\mathbf{m} = \mathbf{M}/M$, where $M$ is a saturation magnetization. Then, $\mathbf{m}_1$ and $\mathbf{m}_2$ are orts along $\mathbf{M}_1$ and $\mathbf{M}_2$: $\mathbf{m}_1 = \mathbf{M}_1/M$, $\mathbf{m}_2 = \mathbf{M}_2/M$. The angle $2\alpha$ between them determines DW type [5] ($2\alpha$-DW): $2\alpha = \arccos(\mathbf{m}_1\mathbf{m}_2)$. The vectors $\mathbf{m}_1$ and $\mathbf{m}_2$ coincide with easy magnetization axes (EMA) and determine boundary conditions of DW. In the crystals with negative CMA the EMAs coincide with <111> like crystallographic directions [5]. The all possible combinations of vectors $\mathbf{m}_1$ and $\mathbf{m}_2$ ($C_8^2$=28 boundary conditions variants) divide into 12, 12 and 4 for the 71°, 109° and 180° DWs respectively.

The orientation of the DW in the unrestricted crystal is defined by the angle $\lambda$ between direction $\mathbf{n}_W$ ($\mathbf{n}_W \perp \Delta\mathbf{m}$ [6]) and plain of vectors $\Delta\mathbf{m}$ and **A**, where $\mathbf{A} = \mathbf{m}_\Sigma /(2\cos\alpha)$ and $\mathbf{A} = [\Delta\mathbf{m} \times \mathbf{P}]/2$ for $71^0$ or $109^0$ DW (fig.1a) and 180° DW (fig.1b) respectively, where $\Delta\mathbf{m} = \mathbf{m}_2 - \mathbf{m}_1$, $\mathbf{m}_\Sigma = \mathbf{m}_1 + \mathbf{m}_2$ and **P** is an ort satisfied the condition $\mathbf{P}\Delta\mathbf{m} = 0$. The **P** is useful to orient along any <110> like direction. In this case and only in this case, the vector **P** is collinear with two fold rotation axis and plate of vectors $\Delta\mathbf{m}$ and **A** coincides with the mirror symmetry plain of the cubic m$\bar{3}$m crystal. The direction of vectors $\Delta\mathbf{m}$ and **A** determine unambiguously by the orientation of vectors $\mathbf{m}_1$ and $\mathbf{m}_2$. With exception of limit values, all possible orientations of DW plain can be unambiguously described in such ranges of $\lambda$ as $-\pi/2 \le \lambda \le \pi/2$ [11]. The interrelation between orientation of the $\mathbf{n}_W$ and the angle $\lambda$ is given by $\mathbf{n}_W = \mathbf{A}\cos\lambda + \mathbf{B}\sin\lambda$, where $\mathbf{B} = \mathbf{P}$ or $\mathbf{B} = [\mathbf{m}_1 \times \mathbf{m}_2]/\sin 2\alpha$ for 180° DW or $71^0$ and $109^0$ DW respectively.

In coordinate system connected with DW plain [11] $O\tilde{x}\tilde{y}\tilde{z}$ ($[\mathbf{e}_{\tilde{x}}, \mathbf{e}_{\tilde{y}}, \mathbf{e}_{\tilde{z}}]$= $[\Delta\mathbf{m} \times \mathbf{n}_W /(2\sin\alpha), \Delta\mathbf{m}/(2\sin\alpha), \mathbf{n}_W]$) the direction **m** is determined by the polar $\tilde{\theta}$ and the azimuth $\tilde{\varphi}$ angles, which are counted from the $O\tilde{z}$ and $O\tilde{x}$ axis respectively (fig.1a-b). Let us set the condition $\mathbf{m}(\tilde{z} \to -\infty) \to \mathbf{m}_1$ and $\mathbf{m}(\tilde{z} \to +\infty) \to \mathbf{m}_2$, that do not restrict model. The spatial distribution of **m** in volume of DW is described by the variable $\tilde{\varphi}$ (value $\tilde{\theta}$ remained constant in the volume of Bloch DW: $\tilde{\theta} = (\mathbf{m}\,\mathbf{n}_W) = \arccos\vartheta$, where $\vartheta = \cos\lambda\cos\alpha$). At $\vartheta = 0$ the rotation of **m** realizes in the plain of the DW. Such rotation of the **m** is solely possible for 180° DW. Also, it realizes in the limit cases ($\lambda = \pm\pi/2$) for 71° and 109° DW.

There are possible opposite directions of **m** rotation between vectors $\mathbf{m}_1$ and $\mathbf{m}_2$ in volume of the DW. The corresponding ranges of $\tilde{\varphi}$ change are bounded by the limit values $\tilde{\varphi}_1 = \tilde{\varphi}(\tilde{z} \to -\infty)$ and $\tilde{\varphi}_2 = \tilde{\varphi}(\tilde{z} \to +\infty)$, where $\tilde{\varphi}_1 = -\tilde{\varphi}_0 + (C_R + 1)\pi$, $\tilde{\varphi}_2 = \tilde{\varphi}_0$, $\tilde{\varphi}_0 = \arccos\left(\cos\alpha \sin\lambda / \sin\tilde{\theta}\right)$, $\alpha < \tilde{\varphi}_0 < \pi - \alpha$. The value $C_R$ is given by $C_R = -1$ and $1$ for right- (*R*-) and left-handed (*L*-) direction of **m** rotation respectively. There are short ($\Delta\tilde{\varphi} < \pi$), middle ($\Delta\tilde{\varphi} = \pi$) and long ($\Delta\tilde{\varphi} > \pi$) paths (*S*-, *M*- and *L*-path respectively) of **m** turn in the volume of the DW, where $\Delta\tilde{\varphi} = |\tilde{\varphi}_2 - \tilde{\varphi}_1|$ [11]. For the $71^0$ and $109^0$ DW value of $\Delta\tilde{\varphi}$ changes in the range $2\alpha \le \Delta\tilde{\varphi} < \pi$ at $\pi/2 \ge \tilde{\lambda} > 0$ and the range $\pi < \Delta\tilde{\varphi} \le 2(\pi - \alpha)$ at $0 > \tilde{\lambda} \ge -\pi/2$ for *S*- and *L*-path respectively, where $\tilde{\lambda} = -C_R\lambda$. The *M*-path is the path of **m** turn in the whole range of changes of $\lambda$ ($-\pi/2 \le \lambda \le \pi/2$) and only in point $\lambda = 0$ for 180° DW and non-180° DW respectively. The kind of path is determined by the parameter $C_P = C_R \,\mathrm{sgn}(\lambda \cos\alpha)$: $C_P = -1$, $C_P = 0$ or $C_P = 1$ for *S*-, *M*- or *L*-path respectively.

Analysis of the above-mentioned paths of **m** rotation leads to the formulation of the following conclusions about spatial position of this rotation. At the $0 < \tilde{\lambda} \le \pi/2$ the **m** in the volume of the 71° DW with *S*-path paths passes two octants of Cartesian coordinate system $Oxyz$ (based on [100], [010] and [001] directions) which contain $\mathbf{m}_1$ and $\mathbf{m}_2$. In the volume of the 109° DW with $S$-path the vector **m** passes two or three octants at the $\tilde{\lambda} = \pi/2$ or $0 < \tilde{\lambda} < \pi/2$ respectively. At the *M*-path in the volume of the 71° or 109° DW the rotation of **m** passes two or three octants respectively. With exceptions of fixed orientations, the rotation of **m** in the volume of the 180° DW passes four octants in the whole region $-\pi/2 \le \lambda \le \pi/2$. The above-mentioned exceptions are the orientations of 180° DW with $\lambda = \pi(d/3 - 1/2)$, $d = \overline{0,3}$, where rotation of **m** passes three octants. In the case of 71° DW with the *L*-path, the rotation of **m** in its volume passes two, four, six and four octants at the $0 \ge \tilde{\lambda} \ge \mu_1$, $\mu_1 > \tilde{\lambda} \ge \mu_2$, $\mu_2 > \tilde{\lambda} > -\pi/2$ and $\tilde{\lambda} = -\pi/2$ respectively. The rotation of **m** in the volume of the 109° DW with *L*-path passes three, four, five or four octants at the $0 \ge \tilde{\lambda} \ge \eta_1$, $\eta_1 > \tilde{\lambda} \ge \eta_2$, $\eta_2 > \tilde{\lambda} > -\pi/2$ or $\tilde{\lambda} = -\pi/2$ respectively. These values are the following: $\mu_1 = -\arccos\sqrt{3/\left[2(5-2\sqrt{3})\right]} \approx -8°48'$, $\mu_2 = -\arccos\sqrt{3/\left[2(5+2\sqrt{3})\right]} \approx -65°6'$, $\eta_1 = -\pi/6 = -30°$ and $\eta_2 = -\arctan\sqrt{2/3} \approx -39°14'$.

### 3. Energy and structure of the domain walls

The equilibrium orientation of the DW in the volumetric plate of the magnetically ordered medium depends on the mutual orientation of the DW plain and the sample surface and corresponds to the minimum of the energy density $\sigma_S = \sigma S/S_0 = \sigma/\sin\psi$, where $\sigma = 2\left|\int_{\tilde{\varphi}_1}^{\tilde{\varphi}_2}\left\{\left[A\sin^2\tilde{\theta}\right]\cdot\left[e_A(\tilde{\theta},\tilde{\varphi}) - e_A(\tilde{\theta},\tilde{\varphi}_1)\right]\right\}^{1/2} d\tilde{\varphi}\right|$ is the energy density of the DW in the unrestricted crystal [5], $A$ is an exchange constant, $e_A(\tilde{\theta},\tilde{\varphi})$ is the CMA energy density, $e_A(\tilde{\theta},\tilde{\varphi}_1) = K_1/3$ [5,6], $\psi$ is an angle between plain of the DW and the surface of the ($nml$) plate, $S_0$ and $S$ is the area of the DW at the ($\mathbf{n}_W\,\mathbf{n}_S$)=0 and arbitrary orientation of the $\mathbf{n}_W$ respectively, $\mathbf{n}_S$ is an ort along the normal of the sample surface: $\mathbf{n}_S = (\mathbf{e}_1 n + \mathbf{e}_2 m + \mathbf{e}_3 l)/\sqrt{u}$, $u = n^2 + m^2 + l^2$, $n$, $m$ and $l$ are the Miller indexes of the sample surface; $\mathbf{e}_1$, $\mathbf{e}_2$ and $\mathbf{e}_3$ are the orts along the [100], [010] and [001] directions respectively. In the coordinate system $O\tilde{x}\tilde{y}\tilde{z}$ the CMA energy density $e_A = K_1\left(\alpha_1^2\alpha_2^2 + \alpha_2^2\alpha_3^2 + \alpha_1^2\alpha_3^2\right)$ is given by expressions:

$$e_A(\tilde{\theta},\tilde{\varphi})/|K_1| = \left[\Pi + (1-\Pi)\sin^2(\Lambda+\pi/4)\sin^2(\Lambda-\pi/4)\right](\Pi-1) \quad \text{for} \quad 71°\text{ DW}, \qquad (1a)$$

$$e_A(\tilde{\theta},\tilde{\varphi})/|K_1| = -(1-\Pi)\cos^2\Lambda\left[1-(1-\Pi)\cos^2\Lambda\right] - \left[\Pi-(1-\Pi)\sin^2\Lambda\right]^2/4 \text{ for } 109°\text{ DW}, \qquad (1b)$$

$$e_A(\tilde{\theta},\tilde{\varphi})/|K_1| = \cos^2\tilde{\varphi}\left(9 - 7\cos 2\tilde{\varphi} - 4\sqrt{2}\sin 2\tilde{\varphi}\sin 3\lambda\right)/24 - 1/3 \text{ for } 180°\text{ DW}, \qquad (1c)$$

where $\Pi = \sin^2\tilde{\theta}\sin^2\tilde{\varphi}$, $\Lambda = \lambda - \lim_{\tau\to 0+}\arctan\left[\cos\tilde{\varphi}\,\tan(\tilde{\theta}-\tau)\right]$, $\alpha_1$, $\alpha_2$ and $\alpha_3$ — are direction cosines of the $\mathbf{m}$ in the $Oxyz$. The following equalities: $e_A(\tilde{\theta},\tilde{\varphi}) = e_A(\tilde{\theta},-\tilde{\varphi})$ and $e_A(\tilde{\theta},\tilde{\varphi}) = e_A(\tilde{\theta},\pi-\tilde{\varphi})$ take place at the any value of $\lambda$ or at the $\lambda \to -\lambda$ respectively. Based on the last expression, the functions $\sigma(\lambda)$ of the semi-type DW with opposite directions of the $\mathbf{m}$ rotation are interrelated by the $\lambda \to -\lambda$. For the 180°

DW the value of $e_A(\tilde{\theta},\tilde{\varphi})$ remains constant at the changing of the $\lambda$ by the angle $\pi/3$ as well as $e_A(\tilde{\theta},\tilde{\varphi})=e_A(\tilde{\theta},\pi-\tilde{\varphi})$ takes place for the any value of $\lambda$. Consequently, values of $\sigma(\lambda)$ for 180° DWs with opposite directions of the **m** rotation are equal.

The function $S(\lambda)$ has parameters $\beta=\arccos|\mathbf{n}_S\mathbf{A}|$, $\gamma=\arccos|\mathbf{n}_S\Delta\mathbf{m}/(2\sin\alpha)|$ and $\nu=\mathrm{sgn}[(\mathbf{A}\mathbf{n}_S)(\mathbf{B}\,\mathbf{n}_S)]$:

$$S(\lambda)=S_0/\sqrt{1-\sin^2\gamma\cos^2[(\nu+|\nu|-1)\lambda-b]}, \qquad (2)$$

where $b=\lim\limits_{\tau\to 0+}\arctan\left[\sqrt{(-\cos 2\beta-\cos 2\gamma)/2}\big/\cos(\beta-\tau)\right]$. The parameters $\beta$ and $\gamma$ (each of them is determined in the region $[0,\pi/2])$ are interrelated: $\cos 2\beta+\cos 2\gamma+\cos 2\phi=-1$, where $\phi=\arccos|\mathbf{n}_S\mathbf{B}|$. The condition for $\phi$ to have real value restricts a totality of possible parameters: $-2\le\cos 2\beta+\cos 2\gamma\le 0$ (fig.2). Internal part of this region is determined by the $0<b<\pi/2$. The value $\nu$ =0 is met at the $\beta=\pi/2$ and/or $\phi=\pi/2$, i.e. at the bound of the region where parameters $\beta$ and $\gamma$ are defined. Geometrical sense of the $\beta$ and $\gamma$ is that they are angles of reflections of vectors **A** and $\Delta\mathbf{m}$ accordingly from the sample plate normal. The angle $\phi$ is an interfacial angle between the sample plain and plain of the vectors **A** and $\Delta\mathbf{m}$ (fig.1c). As it was mentioned in the initial definitions our model is invariant when vectors $\mathbf{m}_1$ and $\mathbf{m}_2$ are transposed. Here, all values and functions are identical with exceptions of value $\nu$ sign inversion. For the 180° DW the values of the $\beta$ and $\phi$, and else $\nu$ depend on choice of the vector **P**. Ambiguous manner of it choice is determined by the six possible <110> like directions and corresponding six functions $\sigma_S=\sigma_S(\lambda)$. These dependence are differed by the shift $k\pi/3$, where $k$=1, 2, 3 taking into account looping at the boundaries of argument region. Unambiguous description of the 180° DW requires the following definition of the **P** choice:

$$\mathbf{P}=\begin{cases}\mathbf{P}_1=s_1(-s_3\mathbf{e}_2+s_2\mathbf{e}_3)/\sqrt{2} & \text{at } |m|<|n|<|l| \text{ or } |m|>|n|>|l| \text{ or } |l|=|m|\neq|n| \\ \mathbf{P}_2=s_2(-s_1\mathbf{e}_3+s_3\mathbf{e}_1)/\sqrt{2} & \text{at } |l|<|m|<|n| \text{ or } |l|>|m|>|n| \text{ or } |n|=|l|\neq|m| \\ \mathbf{P}_3=s_3(-s_2\mathbf{e}_1+s_1\mathbf{e}_2)/\sqrt{2} & \text{at } |n|<|l|<|m| \text{ or } |n|>|l|>|m| \text{ or } |m|=|n|\neq|l| \\ \mathbf{P}_k\ (k=1,2\text{ or }3) & \text{at } |n|=|m|=|l| \text{ and } \gamma=0 \\ \mathbf{P}_k\perp\mathbf{n}_S & \text{at } |n|=|m|=|l| \text{ and } \gamma\neq0\end{cases} \quad (3),$$

where $s_k=\sqrt{3}\Delta\mathbf{m}\mathbf{e}_k/2$ ($k$=1, 2, 3).

The area remains constant $S(\lambda)=S_0$ at the $\gamma=0$. In the other cases (at the $\gamma>0$), dependences $S(\lambda)$ has minimum at the value of $\lambda$, that is equal to the $\lambda_0=(\nu+|\nu|-1)(b-\pi/2)$ or $\lambda_0=\pm\pi/2$ accordingly at the $b>0$ or $b=0$ and maximum value (including limit case $S(\lambda)\to\infty$ at the $\gamma=\pi/2$) when $\lambda$ equals $\lambda_m=(\nu+|\nu|-1)b$ or $\lambda_m=\pm\pi/2$ at the $b<\pi/2$ or $b=\pi/2$ respectively. There are the symmetry $S(\lambda_0+\Delta\lambda)=S(\lambda_0-\Delta\lambda)$ and $S(\lambda_m+\Delta\lambda)=S(\lambda_m-\Delta\lambda)$. Dependences $S(\lambda)$ for DWs with different $\nu$ signs are interrelated by the $\lambda\to-\lambda$. For symmetrical reflections of this DW from the $\lambda$=0 the lower areas $S$ correspond to the ranges containing the $\lambda_0$ (i.e. $-\pi/2<\lambda\nu<0$)**:** $S(\lambda)<S(-\lambda)$. For the $\nu=0$ only the values $\lambda_0$ and $\lambda_m$ equals to the 0 or $\pm\pi/2$.

In general case, dependence $\sigma_S(\lambda)$ in the unique manner is defined by the $\beta$, $\gamma$, and $\nu$ or $\tilde{\nu}$ for the non-180° or 180° DW respectively, where $\tilde{\nu}=\nu C_R$. The rotation direction does not correspond to the specific value of the energy: both *R*- and *L*- rotations are energetically more preferable for different orders of domains (interrelated by the $\mathbf{m}_1\leftrightarrow\mathbf{m}_2$) if all others things are equal. The equilibrium DWs with opposite directions of the **m** rotations have the identical energy density $\sigma_S$ when the function $S(\lambda)$ is even for 71° or 109° DW ($\nu=0$) and always for the 180° DW ($\vartheta=0$). Equilibrium 71° or 109° DW have the rotation of **m** in their plain ($\vartheta=0$) only at the *b*=0 ($\phi=\beta+\gamma=\pi/2$). The general criterion of the coincidence of the equilibrium values $\sigma_S$ of DWs with opposite directions of the **m** rotations is given by $\nu\vartheta=0$. Here, orientations of the equilibrium $71^0$ or $109^0$ DWs with opposite directions of the **m** rotation is given by the angles $\lambda_I=-\lambda_{II}$.

The dependence $\sigma_S(\lambda)$ can coincide with the one $\sigma(\lambda)$ (in the unrestricted crystal) completely at the $\gamma = 0$ (curve 1 on the fig.3a-b) or in the points $\lambda = \lambda_0$ (curve 2 on the fig.3a-b). If the value $\lambda_0$ corresponds to the equilibrium value in the unrestricted crystal then equilibrium DW will be the same too. This case realizes when $\lambda_0 = 39°44' \cdot C_R$ or $\lambda_0 = \pm\pi/2$, and else when $\lambda_0 = -10°9' \cdot C_R$ or $\lambda_0 = -C_R\pi/2$ for $71°$, and else for $109°$ DW respectively.

If $\nu\vartheta \neq 0$ takes place then equilibrium $71^0$ and $109^0$ DWs with opposite directions of the **m** rotations (at $0 < b < \pi/2$) have different equilibrium values $\sigma_S$ (3-4 on the fig.3a-b). If equilibrium DW with specific direction of the rotation presents in the region $-\pi/2 < \lambda\nu < 0$ then it is more preferable then this DW with opposite direction.

The spatial distribution $\tilde{z}$ of the $\tilde{\varphi}$ of the **m** rotation in the volume of the DW is given by the $\tilde{z} = \int\limits_{\varpi}^{\tilde{\varphi}} \left\{ \left[ A\sin^2\tilde{\theta} \right] / \left[ e_A\left(\tilde{\theta}, \tilde{\zeta}\right) - e_A\left(\tilde{\theta}, \tilde{\varphi}_1\right) \right] \right\}^{1/2} d\tilde{\zeta}$, where $\varpi = 0$ and $\pi$ for the *R*- and *L*- rotation in the volume of the DW respectively. The peculiarities of the spatial distribution of the **m** is given by the number of the reflection points of the distribution $\tilde{z}(\tilde{\varphi})$. This number is given by the number of the zero values of the function $\partial e_A\left(\tilde{\theta}, \tilde{\varphi}\right)/\partial\tilde{\varphi}$ in the ranges $\tilde{\varphi}_1 \le \tilde{\varphi} \le \tilde{\varphi}_2$ and $\tilde{\varphi}_2 \le \tilde{\varphi} \le \tilde{\varphi}_1$ for *R*- and *L*-rotations respectively. The **m** distribution in the volume of the 180° DW is always non-monotone (has more than one reflection point) and the number of reflection points $\tilde{\varphi}_k^*$ of the $\tilde{z}(\tilde{\varphi})$ equals three in the range $[-\pi/2 \le \lambda \le \pi/2]$:

$$\tilde{\varphi}_1^* = \varpi + \Xi \sec^{-1}\left[ \sqrt{I\left(I+\sqrt{3}\right)\rho^2 + 3\rho(9 - \cos 6\lambda) + I\left(I-\sqrt{3}\right)\left(4 + 8\sin^2 3\lambda + 9\sin^4 3\lambda\right)} / \left(4\sqrt{\rho}\right) \right] \quad (4a)$$

$$\tilde{\varphi}_2^* = \varpi - \Xi \sec^{-1}\left[ \sqrt{\left(-1 - I\sqrt{3}\right)\rho^2 + 3\rho(9 - \cos 6\lambda) + I\left(I+\sqrt{3}\right)\left(4 + 8\sin^2 3\lambda + 9\sin^4 3\lambda\right)} / \left(4\sqrt{\rho}\right) \right] \quad (4b)$$

$$\tilde{\varphi}_3^* = \varpi + \Xi \sec^{-1}\left[ \sqrt{4 + 12\rho + \rho^2 + (8 + 3\rho)\sin^2 3\lambda + 9\sin^4 3\lambda} / \left(2\sqrt{2\rho}\right) \right] \quad (4c)$$

where $\rho = \left[ \left(26 + 36\sin^2 3\lambda + 27\sin^4 3\lambda\right)\sin^2 3\lambda - 8 + 10\sqrt{-\sin^2 3\lambda\left(8 + 11\sin^2 3\lambda + 8\sin^4 3\lambda\right)} \right]^{1/3}$, $I = \sqrt{-1}$ and $\Xi = \mathrm{sgn}(\sin 3\lambda)$. Here and hereinafter the reflection points $\tilde{\varphi}_2^*$ and $\tilde{\varphi}_3^*$ are the most remote [5] reflection points of the function $\tilde{z}(\tilde{\varphi})$. The expressions (4) give the real values of the $\tilde{\varphi}_k^*$, where $k=\overline{1,3}$. Here and hereinafter all functions of complex argument are token in sense of value principal.

The **m** distribution in the volume of the 71° and 109° DW has the central reflection point $\tilde{\varphi}_1^* = \varpi$ . The roots of the equality $\partial e_A(\tilde{\theta},\tilde{\varphi})/\partial\tilde{\varphi}=0$ in the considering region of $\tilde{\varphi}$ gives the following reflection points. They are given for 71° DW:

$$\tilde{\varphi}_3^* = -\tilde{\varphi}_2^* + 2\varpi = \arccos\left\{-f_1(\tilde{\lambda},\tilde{\theta})/\left[36\left(7-\cos 4\tilde{\lambda}\right)\sin\tilde{\theta}\right]\right\} + \varpi \quad \text{(5a)},$$

$$\tilde{\varphi}_5^* = -\tilde{\varphi}_4^* + 2\varpi = \arccos\left\{-g_1(\tilde{\lambda},\tilde{\theta})/\left[36\left(7-\cos 4\tilde{\lambda}\right)\sin\tilde{\theta}\right]\right\} + \varpi \quad \text{(5b)}$$

and for the 109° DW:

$$\tilde{\varphi}_3^* = -\tilde{\varphi}_2^* + 2\varpi = \min\left\{\arccos\left[f_2(\tilde{\lambda},\tilde{\theta})\right], \arccos\left[g_2(\tilde{\lambda},\tilde{\theta})\right]\right\} + \varpi \quad \text{(6a)},$$

$$\tilde{\varphi}_5^* = -\tilde{\varphi}_4^* + 2\varpi = \arccos\left[f_3(\tilde{\lambda},\tilde{\theta})\right] + \varpi \quad \text{(6b)}.$$

The analytical expressions for the functions $f_1(\tilde{\lambda},\tilde{\theta})$, $f_2(\tilde{\lambda},\tilde{\theta})$, $g_1(\tilde{\lambda},\tilde{\theta})$ and $g_2(\tilde{\lambda},\tilde{\theta})$ are presented in the Appendix A. The range of the orientations of the DW where these reflection points exist is restricted by the condition of real value: $\mathrm{Im}\,\tilde{\varphi}_k^* = 0$, where $k=\overline{1,5}$. For the 71° DW the number of the reflection points equals one and five in the regions $\pi/2 \ge \tilde{\lambda} \ge \tilde{\lambda}_1$ at $\mathrm{Im}\, f_1(\tilde{\lambda},\tilde{\theta}) \ne 0$ and $\tilde{\lambda}_1 > \tilde{\lambda} > -\pi/2$ at $\mathrm{Im}\, f_1(\tilde{\lambda},\tilde{\theta}) = 0$ respectively. For the 109° DW the number of the reflection points equals one, three and five in the regions $\pi/2 \ge \tilde{\lambda} \ge \lambda_2$, $\lambda_2 > \tilde{\lambda} \ge \lambda_3$ and $\lambda_3 > \tilde{\lambda} > -\pi/2$ respectively. The values of the angles $\tilde{\lambda}_2$ and $\tilde{\lambda}_3$ are given by the equations $f_2(\tilde{\lambda},\tilde{\theta}) = 1$ and $f_3(\tilde{\lambda},\tilde{\theta}) = 1$. Analytical expression for the function $f_3(\tilde{\lambda},\tilde{\theta})$ has been presented in the Appendix A. The values of these angles are the following: $\tilde{\lambda}_1 \approx -29.366°$, $\tilde{\lambda}_2 \approx 27.784°$ and $\tilde{\lambda}_3 \approx -66.669°$.

The monotone **m** distribution is symmetrical, i.e. $\tilde{\varphi}(\tilde{z}) - \varpi = -\tilde{\varphi}(-\tilde{z}) + \varpi$ . The **m** distribution is symmetrical in the volumes of the 71° and 109° DW at their arbitrary orientation and else in the volume of the 180° DW at the orientation along the vector $\mathbf{n}_W$ along the directions like <112>, with extra condition $\mathbf{n}_W \perp \Delta\mathbf{m}$: $\pm(s_1 s_3 \mathbf{e}_1 + s_2 s_3 \mathbf{e}_2 - 2\mathbf{e}_3)/\sqrt{6}$, $\pm(s_1 s_2 \mathbf{e}_1 - 2\mathbf{e}_2 + s_3 s_2 \mathbf{e}_3)/\sqrt{6}$ and $\pm(-2\mathbf{e}_1 + s_2 s_1 \mathbf{e}_2 + s_3 s_1 \mathbf{e}_3)/\sqrt{6}$. They correspond to the orientations of the 180° DW with $\lambda = 0$ or $\pm\pi/3$.

### 4. Limit and unrestricted width of the domain walls

It was determined that some Bloch DW orientations correspond to the "infinity" width of the DW [3]. Appearing of the unstable DWs can be the consequence of the applying of the external magnetic field leading to the reorientation of the DW plane and/or reorientation of the domain magnetization directions. So, the general analysis of the DW width is required. This can be obtained analytically based on the expressions (4-6).

The analytical expressions for the width of the DW were obtained for the monotone [5-6] and non-monotone (180° DW only) [3] **m** distribution in the volume of the DW. For the 71° and 109° DWs with monotone **m** distributions the expressions are given by: $\bar{\delta}=2\sqrt{6}L\Big/\sqrt{5-3\cos 4\left(\tilde{\theta}-|\lambda|\right)}$ and $\bar{\delta}=4\sqrt{3}\,L\Big/\left[1+3\cos 2\left(\tilde{\theta}-|\lambda|\right)\right]$, where $L=\Delta\tilde{\varphi}\sin\tilde{\theta}$ is the length of the path of the **m** rotation [11]. Expressions for the 180° DW with non-monotone **m** distribution was presented in [3]. Calculation of the width of DWs with remained types and non-monotone **m** distribution requires determination of the reflection points $\tilde{\varphi}_2^*$ and $\tilde{\varphi}_3^*$. These points were determined by the graphical or numerical methods [3]. Taking into account (5) and (6) we obtained that DW width is given by the expression for $R$ –rotation:

$$\delta=\delta_0\left\{\frac{\left(\tilde{\varphi}_2-\tilde{\varphi}_3^*\right)}{\sqrt{\bar{e}_A\left(\tilde{\theta},\tilde{\varphi}_3^*\right)-\bar{e}_A\left(\tilde{\theta},\tilde{\varphi}_1\right)}}+\frac{\left(\tilde{\varphi}_2^*-\tilde{\varphi}_1\right)}{\sqrt{\bar{e}_A\left(\tilde{\theta},\tilde{\varphi}_2^*\right)-\bar{e}_A\left(\tilde{\theta},\tilde{\varphi}_1\right)}}\right\}\sin\tilde{\theta}+\tilde{z}\left(\tilde{\varphi}_3^*\right)-\tilde{z}\left(\tilde{\varphi}_2^*\right) \qquad (7a)$$

and for the $L$ – rotation:

$$\delta=\delta_0\left\{\frac{\left(\tilde{\varphi}_1-\tilde{\varphi}_3^*\right)}{\sqrt{\bar{e}_A\left(\tilde{\theta},\tilde{\varphi}_3^*\right)-\bar{e}_A\left(\tilde{\theta},\tilde{\varphi}_1\right)}}+\frac{\left(\tilde{\varphi}_2^*-\tilde{\varphi}_2\right)}{\sqrt{\bar{e}_A\left(\tilde{\theta},\tilde{\varphi}_2^*\right)-\bar{e}_A\left(\tilde{\theta},\tilde{\varphi}_1\right)}}\right\}\sin\tilde{\theta}+\tilde{z}\left(\tilde{\varphi}_3^*\right)-\tilde{z}\left(\tilde{\varphi}_2^*\right), \qquad (7b),$$

where $\delta_0=\sqrt{A/K_1}$ and $\bar{e}_A\left(\tilde{\theta},\tilde{\varphi}\right)=e_A\left(\tilde{\theta},\tilde{\varphi}\right)/|K_1|$.

In the case of the non-monotone **m** distribution the function $\tilde{\varphi}(\tilde{z})$ can contain reflection points with $\partial^2 e_A\left(\tilde{\theta},\tilde{\zeta}\right)/\partial\tilde{\zeta}^2>0$. These points lead to the growth (unrestricted in the limit case) of the DW width. In the scope of the current model the unrestricted DW width realizes when **m** passes EMAs. This situation takes place at the critical orientations $\tilde{\lambda} = \tilde{\lambda}_J^C$ of the DW plain: $\mathbf{n}_W\left(-C_R\tilde{\lambda}_J^C\right)\cdot\mathbf{m}_1 = \mathbf{n}_W\left(-C_R\tilde{\lambda}_J^C\right)\cdot\mathbf{u}_J$, where $\mathbf{u}_J$ is the ort along EMA ($\mathbf{u}_J \neq \mathbf{m}_1; \mathbf{u}_J \neq \mathbf{m}_2$), where $J$ – is an integer. There are the following critical orientations: $\tilde{\lambda}_1^C = -\pi/4$ and $\tilde{\lambda}_2^C = -\pi/2$ for the 71° DW; $\tilde{\lambda}_1^C = 0$, $\tilde{\lambda}_2^C = -\arccos\left(1/\sqrt{3}\right)$ and $\tilde{\lambda}_3^C = -\pi/2$ for the 109° DW; $\tilde{\lambda}_1^C = -\pi/2$, $\tilde{\lambda}_2^C = -\pi/6$, $\tilde{\lambda}_3^C = \pi/6$ and $\tilde{\lambda}_4^C = \pi/2$ for the 180° DW. The DWs with such plain orientations are unstable, i.e. they disintegrate when obtain such orientation.

In the such cases the 71° DW divides into three 71° DWs at the $\tilde{\lambda} = \tilde{\lambda}_1^C$ or two 109° DWs and one 71° DW at the $\tilde{\lambda} = \tilde{\lambda}_2^C$. The 109° DW divides into the two 71° DWs at the $\tilde{\lambda} = \tilde{\lambda}_1^C$, two 109° DWs at the $\tilde{\lambda} = \tilde{\lambda}_2^C$ or two 71° DWs and one 109° DW at the $\tilde{\lambda} = \tilde{\lambda}_3^C$. The 180° DW divides into the 71° DW and 109° DW at any critical orientation. The conservation of the length [5,11] of the **m** rotation takes place: $L = \sum_k L_k$ as well as the conservation of the energy density $\sigma_S = \sum_k \sigma_S^k$ and $\Delta\tilde{\varphi} = \sum_k \Delta\tilde{\varphi}_k$, where $L_k$ = $\Delta\tilde{\varphi}_k \sin\tilde{\theta}_k$. Here, $\Delta\tilde{\varphi}_k$ and $\sigma_S^k$ are the corresponding parameters of DWs which born after disintegration of the initial DW (these DWs are numbered by the $k$). All paths of **m** rotation of the initial DWs are $L$- or $M$-paths. All paths of the new DWs are the $S$-paths.

The stability and restrictions of the unstable DWs is leading by the taking into account of the extra components of energy [3]. For example, it can be magnetostriction energy term. Thus, consideration of the DWs in the crystal with small magnetostriction is of interest [21,22]. Also, taking into account demagnetization energy term leads to stabilization of the above-mentioned DWs as well as rising of the new types of unstable DW types [23].

## 5. Schema of the domain walls degeneration in the (*nml*) plates

The above mentioned theory allows to describe and to analyze the energy and the structure (plain orientation, magnetization distribution, DW width) of the 71°, 109° and 180° DW with arbitrary boundary conditions in the plate with arbitrary oriented surface. The analysis of these expressions leads to formulate the main laws for these DWs. Equilibrium DWs in the restricted crystal corresponds to the minimum of the energy density $\sigma_S$ and do not coincide with such DWs in the unrestricted crystal in general case. It leads to dividing of the semi-type equilibrium DWs into the $Q_{2\alpha}$ groups. The semi-types DWs from different groups have different areas and values $\sigma_S$. Each group includes $q$ semi-types DWs with the identical values $\sigma_S$ but different $\mathbf{A}$ and/or $\Delta\mathbf{m}$. The criterion of the equal values $\sigma_S$ of semi-type DWs with specific **m** rotation path spatial position is the equality of their areas. The last is determined by the identical angle pairs $\beta$ and $\gamma$. In general case the values $q$ are different for different groups. In the arbitrary (*nml*) plate two $71^0$ DWs (or two $109^0$ DWs) belong to group if their vectors A (or $\Delta\mathbf{m}$) are collinear to the directions $(\mathbf{e}_i + \mathbf{e}_j)/\sqrt{2}$ and $(\mathbf{e}_i - \mathbf{e}_j)/\sqrt{2}$ (in the (*nml*) plates with $S_i S_j = 0$, where $S_i = (\mathbf{n}_S \mathbf{e}_i)$, $S_j = (\mathbf{n}_S \mathbf{e}_j)$, $i \neq j$, $i, j = 1, 2, 3$), directions $(\mathbf{e}_i \pm \mathbf{e}_j)/\sqrt{2}$ and $(\mathbf{e}_i \pm \mathbf{e}_k)/\sqrt{2}$ (in the (*nml*) plates with $S_i = \pm S_j$ or $S_i = \mp S_j = \mp S_k$ or simultaneous $S_i = 0$ and $\pm S_j = \mp S_k$, where $S_k = (\mathbf{n}_S \mathbf{e}_k)$, $k \neq i, j$, $k$=1,2,3). Two $180^0$ DW belong to the group if their vectors $\Delta\mathbf{m}$ are collinear to the directions $(\mathbf{e}_i \pm \mathbf{e}_j + \mathbf{e}_k)/\sqrt{3}$ and $(-\mathbf{e}_i \pm \mathbf{e}_j + \mathbf{e}_k)/\sqrt{3}$ (in the (*nml*) plates with $S_i = 0$ or $S_j = -S_k$), directions $(\mathbf{e}_i + \mathbf{e}_j - \mathbf{e}_k)/\sqrt{3}$ and $(-\mathbf{e}_i + \mathbf{e}_j - \mathbf{e}_k)/\sqrt{3}$ (in the (*nml*) plates with $S_i = 0$ or $S_j = S_k$) or directions $(\mathbf{e}_i - \mathbf{e}_j - \mathbf{e}_k)/\sqrt{3}$ and $-(\mathbf{e}_i + \mathbf{e}_j + \mathbf{e}_k)/\sqrt{3}$ (in the (*nml*) plates with $S_i = 0$ or $S_j = -S_k$).

All possible orientations of plate are described by the six combination of Miller indexes: (100), (110), (111), (*nnl*), (*nm*0) and (*nml*), where *n*, *m* and *l* are the nonzero values with different magnitudes. Results of application of the above-mentioned criteria to the all possible DWs are summarized in table 1-3. The number of groups $Q_{2\alpha}$ in the (*nml*) plates growths with reflection of the sample surface normal $\mathbf{n}_S$ from the high-symmetric crystallographic directions <001>, <110> and <111>. The minimal number of this groups realizes in the (001) plate.

In the spatially restricted crystal 71° DW corresponds to specific 109° DW: $\beta' = \gamma''$ and $\gamma' = \beta''$, where angles with one and two accents relates to the 71° or 109° DW respectively. Vector **B** of the 71° DW always coincides with the vector **A** of some another 71° DW in the (*nml*) plate: angle $\phi$ of the first DW coincides with the angle $\beta$ of the second one. These 71° DW have collinear vectors $\Delta\mathbf{m}$ and identical angles $\gamma$. There are pairs of semi-types non-180° DWs with the same $\phi$ (and different $\beta$ and/or $\gamma$) in the (*nml*) plates which do not contain <110> like directions and contain <112> like directions. They are (*nm*0) and (*nml*) plates with $D=\left(|n|-|m|\right)\left(|m|-|l|\right)\left(|l|-|n|\right)\neq 0$ and $T=\left(2|m|-|l\pm n|\right)\left(2|n|-|m\pm l|\right)\left(2|l|-|n\pm m|\right)=0$.

The DWs with opposite **m** rotation directions (at the same domain order) but the same energy realize in the samples with *D*=0: (100), (110), (111) and (*nnl*) plates. For the 109° DWs this totality of plates is wider. It contains also plates with <100> like directions in their plane (*nml*=0). When vector **A** of the one 71° DW coincides with the vector **B** of another 71° DW then these DWs can have equal energies at the opposite **m** rotations only simultaneously. The equilibrium 71° and 109° DWs have **m** rotation in their plane in samples with *D*=0. The equilibrium 180° DWs with different orientations of their planes realize in the samples with <110> or <112> like directions in their plain (*TD*=0, $T\neq D$) if they have <111> like direction reflected by the angle 40° from the sample normal.

The equilibrium DWs identical to ones in the unrestricted crystal realize in the samples with *D*=0 (they are DWs with **m** rotation in their planes) and also in the (*nml*) plates satisfied the conditions:

$$\left(10.831|n|-|m|\right)\left(10.831|m|-|l|\right)\left(10.831|l|-|n|\right)\left(10.831|m|-|n|\right)\left(10.831|l|-|m|\right)\left(10.831|n|-|l|\right)=0 \qquad (8a)$$

or

$$\left(1.436|n|-|m|\right)\left(1.436|m|-|l|\right)\left(1.436|l|-|n|\right)\left(1.436|m|-|n|\right)\left(1.436|l|-|m|\right)\left(1.436|n|-|l|\right)=0 \qquad (8b)$$

for the 71° and 109° DW respectively. In the specially oriented (001), (110), (111), (112) and the (210) plates they are only $180^0$ DWs ($\phi=90^0$; $\beta=90^0$; $\gamma=0^0$) in the (111) plate, $109^0$ DW ($\phi=90^0$; $\beta=90^0$; $\gamma=0^0$) in the (110) plate and the $71^0$ DW ($\phi=90^0$; $\beta=90^0$; $\gamma=0^0$) in the (001) plate (curve 1 on the fig.3).

As a rule, there are in the restricted bulk crystals DW orientation changes as well as the energy density growths ($\sigma_S > \sigma$ ) in relation with unrestricted crystals (curves 3-4 on the fig.3). Here, partial or full removing of the semi-type DWs degeneration takes place in the (*nml*) plates.

The equilibrium orientations of the DWs in the above-mentioned plates have been investigated by the numerical integration. Generally, the equilibrium DWs with the *S*-path of **m** rotation have lower values of the $\sigma_S$ as well as the values of width $\delta$. There are several exceptions of this rule. The 71° DW ($\phi=45^0$; $\beta=45^0$; $\gamma=90^0$) and the 109° DW ($\phi=71.565^0$; $\beta=26.565^0$; $\gamma=71.565^0$) with the *R* – rotation and $L$ - path are have the lower energies. They realize in (001) and (210) plate respectively. One more exception is the 109° DW ($\phi=60^0$; $\beta=45^0$; $\gamma=60^0$) with *R* – rotation in the (110) plate. The unique peculiarity of this DW is very small value $|\lambda|$<3.7° at the equilibrium orientation.

The DWs with some received equilibrium orientations were experimentally observed in (110)[9,10], (111) [15,16] and (112) plates of iron yttrium garnet [9,16], nickel [10,13] and $U_3P_4$ single crystal [15]. Their parameters are italicized in the appropriate tables of the Appendix B. The unstable DWs are bold in the Appendix B.

## 6. Conclusions

The typical for the unrestricted crystal degeneration of the semi-type DWs is removed in spatially restricted bulk crystal. The bulk (*nml*) plate contains restrict number of the DWs with different values of the area and the energy. This number is determined by the sample orientation and depends on the magnetization vector orientations of the domains relatively to sample surface. It growth with reflections of sample normal from the high-symmetrical <001>, <110> and <111> like orientations The minimal number of DWs with different energies is realized in the (001) plate.

The most DWs identical to the ones in the unrestricted crystal are possible in the bulk (*nml*) plates with orthogonal planes of domains magnetization directions and the sample surface. This DWs have absolutely minimal values of the areas and energies among the all semi-types DWs.

All DWs with opposite directions of the magnetization rotation are equal energy at the rotation in the DW plane. There are equal energy 71° DWs with opposite directions of the magnetization rotation in cases when <110> directions are parallel with the sample plane. There are also equal energy 109° DWs with opposite directions of the magnetization rotation in cases when <110> or <100> directions are parallel with the sample plane. All 180° DWs with opposite directions of rotation are equal energy in the arbitrary oriented (*nml*) plate.

The spatially restricted sample allows equilibrium 71° and 109° DWs with the long path of magnetization rotation. All typical specially oriented plates allow appearing of the unstable DWs which can be considered as the way of the magnetization reversal of the sample.

**Table 1.** The degeneracy $q$ of the plain Bloch $180°$ DWs in the ($nml$) plates

<table>
<tr><td rowspan="3">$\pm\mathbf{\Delta m}$</td><td colspan="12">$n \ge 0$, $m \ge 0$, $l \ge 0$</td></tr>
<tr><td colspan="2">$n=m=0$; $l \ne 0$</td><td colspan="2">$n=m==l$</td><td colspan="2">$m=l\ne n=0$</td><td colspan="2">$m \ne l$; $m,l \ne 0$; $n=0$</td><td colspan="2">$n=m \ne l \ne 0$; $n \ne 0$</td><td colspan="2">$n \ne m \ne l$; $n \ne l$; $n,m,l \ne 0$</td></tr>
<tr><td>$\gamma$,°</td><td>$q$</td><td>$\gamma$,°</td><td>$q$</td><td>$\gamma$,°</td><td>$q$</td><td>$\gamma$, rad</td><td>$q$</td><td>$\gamma$, rad</td><td>$q$</td><td>$\gamma$, rad</td><td>$q$</td></tr>
<tr><td>$[111]$</td><td rowspan="4">54.74</td><td rowspan="4">4</td><td>0</td><td>1</td><td rowspan="2">35.26</td><td rowspan="2">2</td><td rowspan="2">$\arccos\left[\frac{|m+l|}{\sqrt{3u}}\right]$</td><td rowspan="2">2</td><td>$\arccos\left[\frac{|l+2n|}{\sqrt{3u}}\right]$</td><td>1</td><td>$\arccos\left[\frac{|n+m+l|}{\sqrt{3u}}\right]$</td><td>1</td></tr>
<tr><td>$[\bar{1}11]$</td><td rowspan="3">71</td><td rowspan="3">3</td><td rowspan="2">$\arccos\left[\frac{|l|}{\sqrt{3u}}\right]$</td><td rowspan="2">2</td><td>$\arccos\left[\frac{|m+l-n|}{\sqrt{3u}}\right]$</td><td>1</td></tr>
<tr><td>$[1\bar{1}1]$</td><td rowspan="2">90</td><td rowspan="2">2</td><td rowspan="2">$\arccos\left[\frac{|m-l|}{\sqrt{3u}}\right]$</td><td rowspan="2">2</td><td>$\arccos\left[\frac{|n-m+l|}{\sqrt{3u}}\right]$</td><td>1</td></tr>
<tr><td>$[11\bar{1}]$</td><td>$\arccos\left[\frac{|l-2n|}{\sqrt{3u}}\right]$</td><td>1</td><td>$\arccos\left[\frac{|n+m-l|}{\sqrt{3u}}\right]$</td><td>1</td></tr>
<tr><td>$Q_{180}$</td><td colspan="2">**1**</td><td colspan="2">**2**</td><td colspan="2">**2**</td><td colspan="2">**2**</td><td colspan="2">**3**</td><td colspan="2">**4**</td></tr>
</table>

**Table 2.** The degeneracy $q$ of the plain Bloch $71°$ and $109°$ DWs in the ($nml$) plates containing the <100> like directions in the plain ($nml$).

<table>
<tr><td rowspan="3">$\pm\mathbf{m}_{\Sigma}(\pm\mathbf{\Delta m})$ for $2\alpha=71°$ or $\pm\mathbf{\Delta m}(\pm\mathbf{m}_{\Sigma})$ for $2\alpha=109°$</td><td colspan="6">$n \ge 0$, $m \ge 0$, $l \ge 0$</td></tr>
<tr><td colspan="2">$n=m=0$; $l \ne 0$</td><td colspan="2">$n=m \ne l=0$</td><td colspan="2">$n \ne m$; $n,m \ne 0$; $l=0$</td></tr>
<tr><td>$\beta\{\gamma\}$,° for $2\alpha=71°$ or $\gamma\{\beta\}$,° for $2\alpha=109°$</td><td>$q$</td><td>$\beta\{\gamma\}$,° for $2\alpha=71°$ or $\gamma\{\beta\}$,° for $2\alpha=109°$</td><td>$q$</td><td>$\beta\{\gamma\}$,rad for $2\alpha=71°$ or $\gamma\{\beta\}$,rad for $2\alpha=109°$</td><td>$q$</td></tr>
<tr><td>$[110]([001])$</td><td rowspan="2">90{0}</td><td rowspan="2">4</td><td>0{90}</td><td>2</td><td>$\arccos\left[\frac{|n+m|}{\sqrt{2u}}\right]\{\pi/2\}$</td><td>2</td></tr>
<tr><td>$[1\bar{1}0]([001])$</td><td>90{90}</td><td>2</td><td>$\arccos\left[\frac{|n-m|}{\sqrt{2u}}\right]\{\pi/2\}$</td><td>2</td></tr>
<tr><td>$[011]([100])$</td><td rowspan="4">45{90}</td><td rowspan="4">8</td><td rowspan="4">60{45}</td><td rowspan="4">8</td><td rowspan="2">$\arccos\left[|m|/\sqrt{2u}\right]$ $\{\arccos\left[|n|/\sqrt{u}\right]\}$</td><td rowspan="2">4</td></tr>
<tr><td>$[01\bar{1}]([100])$</td></tr>
<tr><td>$[101]([010])$</td><td rowspan="2">$\arccos\left[|n|/\sqrt{2u}\right]$ $\{\arccos\left[|m|/\sqrt{u}\right]\}$</td><td rowspan="2">4</td></tr>
<tr><td>$[\bar{1}01]([010])$</td></tr>
<tr><td>$Q_{71}$ or $Q_{109}$</td><td colspan="2">**2**</td><td colspan="2">**3**</td><td colspan="2">**4**</td></tr>
</table>

**Table 3.** The degeneracy $q$ of the plain Bloch $71°$ and $109°$ DWs in the (*nml*) plates not containing the <100> like directions in the plain (*nml*).

| $\pm\mathbf{m}_\Sigma(\pm\Delta\mathbf{m})$ for $2\alpha$=71° or $\pm\Delta\mathbf{m}(\pm\mathbf{m}_\Sigma)$ for $2\alpha$=109° | $n\geq0,\ m\geq0,\ l\geq0$ | | | | | |
|---|---|---|---|---|---|---|
| | $n$=$m$=$l$ | | $n=m\neq l\neq 0;\ n\neq 0$ | | $n\neq m\neq l;\ n\neq l;$ $n,m,l\neq 0$ | |
| | $\beta\{\gamma\}$,° for $2\alpha$=71° or $\gamma\{\beta\}$,° for $2\alpha$=109° | $q$ | $\beta\{\gamma\}$,rad for $2\alpha$=71° or $\gamma\{\beta\}$,rad for $2\alpha$=109° | $q$ | $\beta\{\gamma\}$,rad for $2\alpha$=71° or $\gamma\{\beta\}$,rad for $2\alpha$=109° | $q$ |
| $[110]([001])$ | 35.26 {54.74} | 6 | $\arccos[\lvert n\rvert\sqrt{2}/\sqrt{u}]$ $\{\arccos[\lvert l\rvert/\sqrt{u}]\}$ | 2 | $\arccos[\lvert n+m\rvert/\sqrt{2u}]$ $\{\arccos[\lvert l\rvert/\sqrt{u}]\}$ | 2 |
| $[101]([010])$ | | | $\arccos[\lvert n+l\rvert/\sqrt{2u}]$ $\{\arccos[\lvert n\rvert/\sqrt{u}]\}$ | 4 | $\arccos[\lvert n+l\rvert/\sqrt{2u}]$ $\{\arccos[\lvert m\rvert/\sqrt{u}]\}$ | 2 |
| $[011]([100])$ | | | | | $\arccos[\lvert m+l\rvert/\sqrt{2u}]$ $\{\arccos[\lvert n\rvert/\sqrt{u}]\}$ | 2 |
| $[\bar{1}10]([001])$ | 90 {54.74} | 6 | $\pi/2$ $\{\arccos[\lvert l\rvert/\sqrt{u}]\}$ | 2 | $\arccos[\lvert n-m\rvert/\sqrt{2u}]$ $\{\arccos[\lvert l\rvert/\sqrt{u}]\}$ | 2 |
| $[0\bar{1}1]([100])$ | | | $\arccos[\lvert n-l\rvert/\sqrt{2u}]$ $\{\arccos[\lvert n\rvert/\sqrt{u}]\}$ | 4 | $\arccos[\lvert m-l\rvert/\sqrt{2u}]$ $\{\arccos[\lvert n\rvert/\sqrt{u}]\}$ | 2 |
| $[10\bar{1}]([010])$ | | | | | $\arccos[\lvert n-l\rvert/\sqrt{2u}]$ $\{\arccos[\lvert m\rvert/\sqrt{u}]\}$ | 2 |
| $Q_{71}$ or $Q_{109}$ | **2** | | **4** | | **6** | |

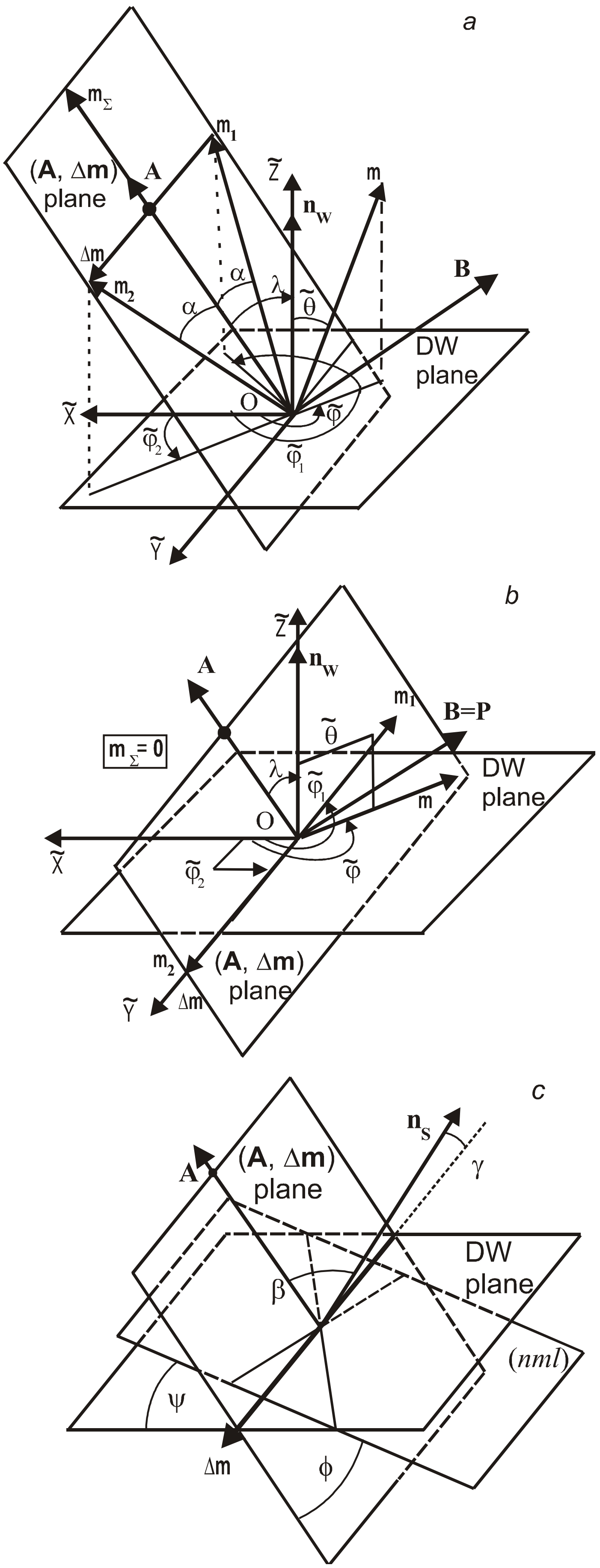

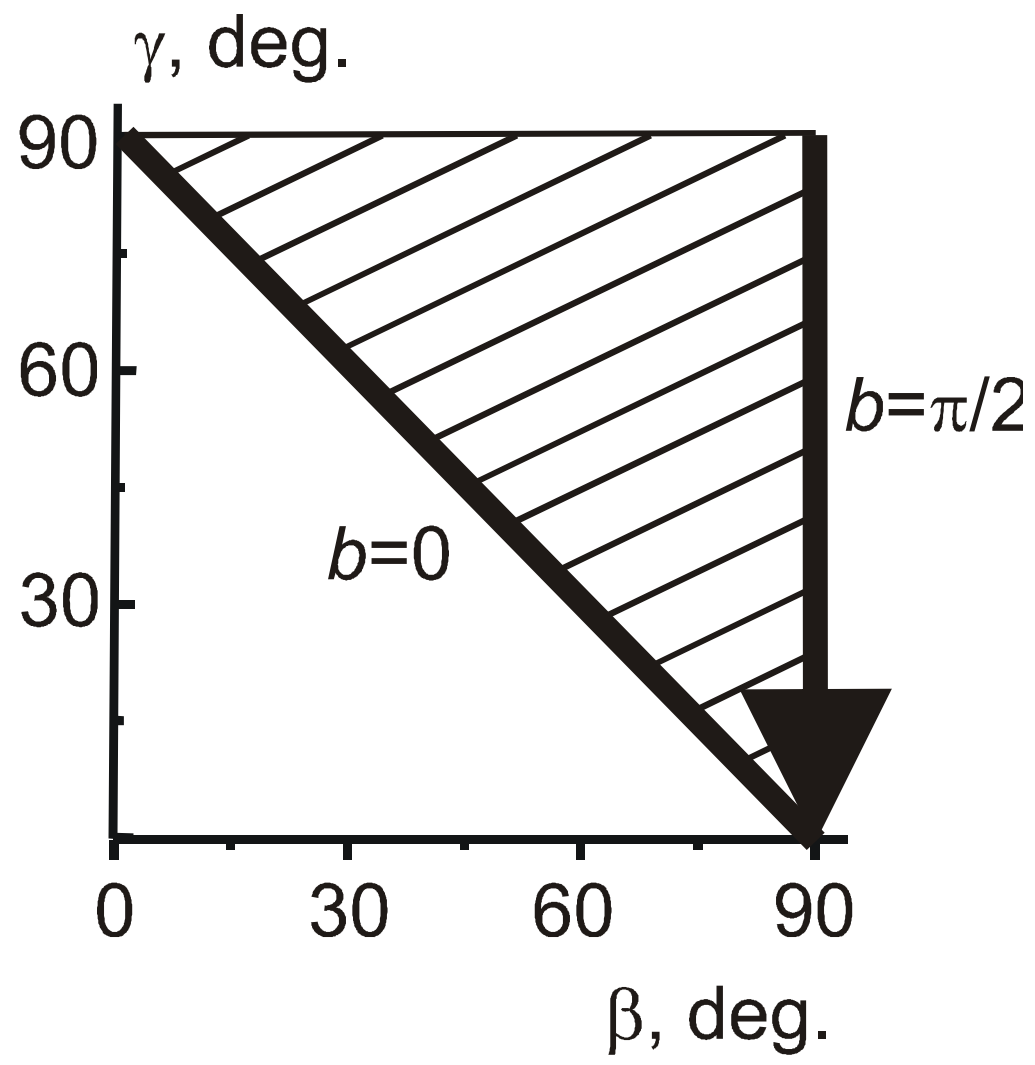

Fig.1. The coordinate system for the $180^0$ DW (a) and non-$180^0$ DW (b); the definition of the angles describing the mutual position of the $\mathbf{m}_1$, $\mathbf{m}_2$, DW plane and sample (*nml*) plane.

Fig.2. The area of possible angles $\beta$ and $\gamma$. The bold line mark special values were $\nu = 0$ .

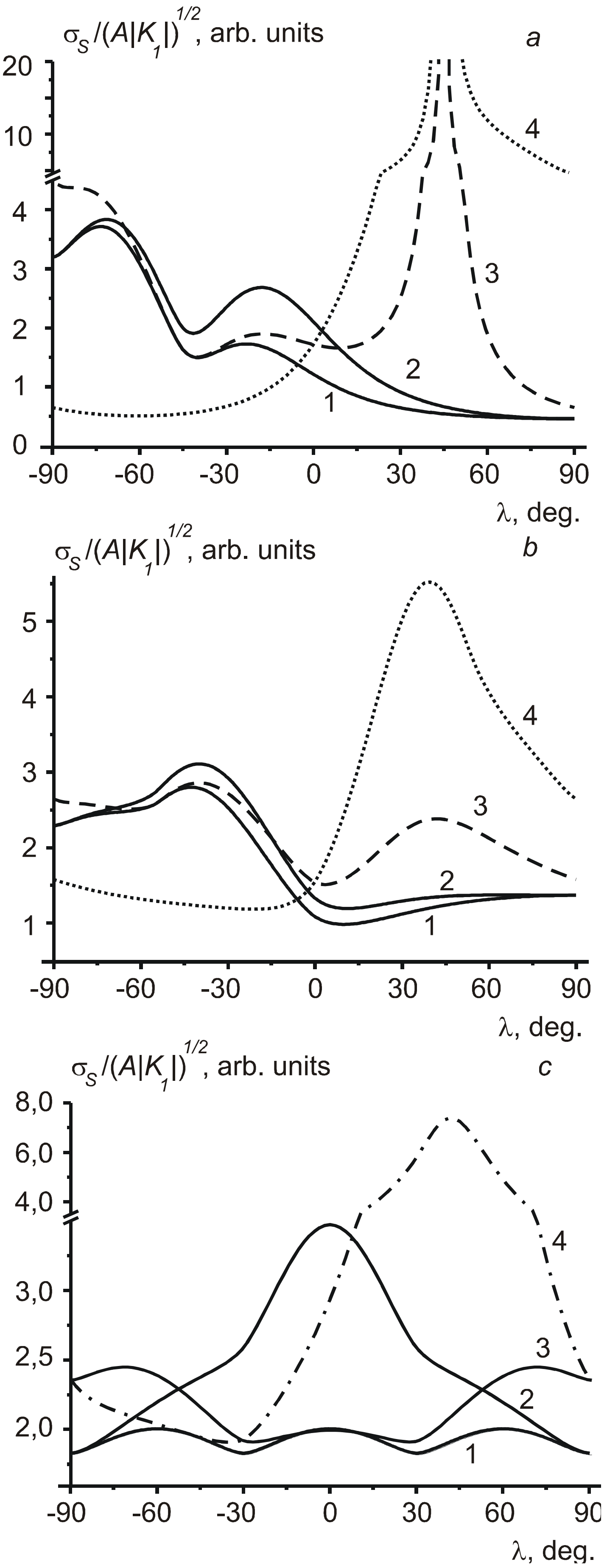

Fig.3. The examples of the orientation dependences of the $71^0$ DW (a), $109^0$ DW (b) and the $180^0$ DW (c):

**a).** The orientation dependences of the energy density of the 71 degree DW with right-handed (1-3) and the left-handed (4) magnetization rotation. The parameters $(\nu, \beta, \gamma)$ are: $\nu = 0$ (curves 1-2) and $\nu = 1$ (curves 3-4); $\beta = 90\,\deg.$ and $\gamma = 0\,\deg.$ (curve 1), $\beta \approx 35.264\,\deg.$ and $\gamma \approx 54.735\,\deg.$ (curve 2), and else $\beta = 45\,\deg.$ and $\gamma = 90\,\deg.$ (curves 3-4).

**b).** The orientation dependences of the energy density of the 109 degree DW with right-handed (1-3) and the left-handed (4) magnetization rotation. The parameters $(\nu, \beta, \gamma)$ are: $\nu = 0$ (curves 1-2) and $\nu = 1$ (curves 3-4); $\beta = 90\,\deg.$ and $\gamma = 0\,\deg.$ (curve 1), $\beta \approx 54.735\,\deg.$ and $\gamma \approx 35.264\,\deg.$ (curve 2), and else $\beta = 45\,\deg.$ and $\gamma = 60\,\deg.$ (curve 3-4).

**c).** The orientation dependences of the energy density of the 109 degree DW with arbitrary direction of the magnetization rotation. The parameters $(\nu, \beta, \gamma)$ are: $\nu = 0$ (curves 1-3) and $\nu = 1$ (curve 4); $\beta = 90\,\deg.$ and $\gamma = 0\,\deg.$ (curve 1), $\beta \approx 35.264\,\deg.$ and $\gamma \approx 54.736\,\deg.$ (curve 2), $\beta = 90\,\deg.$ and $\gamma \approx 39.232\,\deg.$ (curve 3), and else $\beta \approx 43.089\,\deg.$ and $\gamma \approx 75.037\,\deg.$ (curve 4).

**Appendix A.**

$$f_1(\tilde{\lambda},\tilde{\theta}) = -162^{1/3}(1+I\sqrt{3})t_1 + 2(18\sqrt{3})^{2/3}(1-I\sqrt{3})[11\cos 4\tilde{\lambda}+\cos 2\tilde{\theta}(23+7\cos 4\tilde{\lambda})-5]/t_1 +$$

$$+36\cos\tilde{\theta}\sin 4\tilde{\lambda}\ ,$$

$$f_2(\tilde{\lambda},\tilde{\theta}) = \csc\tilde{\theta}\{36\cos\tilde{\theta}(3\sin 4\tilde{\lambda}-14\sin 2\tilde{\lambda})+[6^{5/3}(I\sqrt{3}+1)(-166+23\cos 2\tilde{\lambda}+6\cos 4\tilde{\lambda}+$$

$$+4\cos 2\tilde{\theta}(28\cos 2\tilde{\lambda}+21\cos 4\tilde{\lambda}-17)+9\cos 6\tilde{\lambda})]/\sqrt[3]{t_2}+(I\sqrt{3}-1)\sqrt[3]{6^4 t_2}\}/[72(3\cos^2 2\tilde{\lambda}-5-14\cos 2\tilde{\lambda})],$$

$$f_3(\tilde{\lambda},\tilde{\theta}) = \csc\tilde{\theta}\{18\cos\tilde{\theta}(3\sin 4\tilde{\lambda}-14\sin 2\tilde{\lambda})-[6^{5/3}(23\cos 2\tilde{\lambda}+6\cos 4\tilde{\lambda}+$$

$$+4\cos 2\tilde{\theta}(28\cos 2\tilde{\lambda}+21\cos 4\tilde{\lambda}-17)+9\cos 6\tilde{\lambda}-166)]/\sqrt[3]{t_2}+\sqrt[3]{6^4 t_2}\}/[36(3\cos^2 2\tilde{\lambda}-5-14\cos 2\tilde{\lambda})],$$

$$g_1(\tilde{\lambda},\tilde{\theta}) = -162^{1/3}(1-I\sqrt{3})t_1 + 2(18\sqrt{3})^{2/3}(1+I\sqrt{3})[11\cos 4\tilde{\lambda}+\cos 2\tilde{\theta}(23+7\cos 4\tilde{\lambda})-5]/t_1 +$$

$$+36\cos\tilde{\theta}\sin 4\tilde{\lambda}\ ,$$

$$g_2(\tilde{\lambda},\tilde{\theta}) = \csc\tilde{\theta}\{36\cos\tilde{\theta}(3\sin 4\tilde{\lambda}-14\sin 2\tilde{\lambda})+[6^{5/3}(1-I\sqrt{3})(-166+23\cos 2\tilde{\lambda}+6\cos 4\tilde{\lambda}+$$

$$+4\cos 2\tilde{\theta}(28\cos 2\tilde{\lambda}+21\cos 4\tilde{\lambda}-17)+9\cos 6\tilde{\lambda})]/\sqrt[3]{t_2}-\sqrt[3]{6^4 t_2}(1+I\sqrt{3})\}/[72(3\cos^2 2\tilde{\lambda}-5-14\cos 2\tilde{\lambda})],$$

where $t_1=\{\sqrt{6}[2(\cos 4\tilde{\lambda}-7)^2(250+1893\cos 2\tilde{\theta}+258\cos 4\tilde{\theta}+407\cos 6\tilde{\theta}+(782+825\cos 2\tilde{\theta}+642\cos 4\tilde{\theta}+$

$$+343\cos 6\tilde{\theta})\cos 4\tilde{\lambda}-216\cos^2\tilde{\theta}\cos 8\tilde{\lambda})]^{1/2}-72(1+8\cos 2\tilde{\lambda}+\cos 4\tilde{\lambda})\sin 4\tilde{\lambda}\cos\tilde{\theta}\}^{1/3},$$

$$t_2=\{-6(7+28\cos 2\tilde{\lambda}-3\cos 4\tilde{\lambda})^2[6394+804\cos 2\tilde{\theta}-4080\cos 4\tilde{\theta}-2416\cos 6\tilde{\theta}-2\cos 2\tilde{\lambda}(4213+$$

$$+8112\cos 2\tilde{\theta}+1140\cos 4\tilde{\theta}+224\cos 6\tilde{\theta})+24\cos 4\tilde{\lambda}(181+228\cos 2\tilde{\theta}+330\cos 4\tilde{\theta}+98\cos 6\tilde{\theta})+$$

$$+27(97+96\cos 2\tilde{\theta}+56\cos 4\tilde{\theta})\cos 6\tilde{\lambda}-54(6\cos 2\tilde{\theta}-5)\cos 8\tilde{\lambda}-81\cos 10\tilde{\lambda}]\}^{1/2}+36[-4\cos 3\tilde{\theta}\times$$

$$\times(35\sin 2\tilde{\lambda}+4\sin 4\tilde{\lambda}+7\sin 6\tilde{\lambda})-\cos\tilde{\theta}(302\sin 2\tilde{\lambda}-50\sin 4\tilde{\lambda}+6\sin 6\tilde{\lambda}+9\sin 8\tilde{\lambda})].$$

## Appendix B. Energy density and structure of the plane Bloch DW (equilibrium states in regular and unstable in bold) in the (*nml*)-plates of crystal with “negative” CMA

Table B.1. Energy density and structure of the plane Bloch DW in (001)-plate

| Int.angl. | DW type | | | **M** turn type | | DW energy density and structure | | | | |
|---|---|---|---|---|---|---|---|---|---|---|
| $\phi$, deg. | $2\alpha$, deg. | $\beta$, deg. | $\gamma$, deg. | Rotation | Path | $\lambda$, deg. | $\Delta\tilde{\varphi}$, deg. | $\psi$, deg. | $\bar{\delta}$, arb.units. | $\bar{\sigma}/\sin\psi$, arb.units. |
| 45.000 | 71 | 45.000 | 90.000 | *R-* | *S-* | 8.923 | 155.257 | 36.078 | 3.206 | 1.659 |
| | | | | | *L-* | -40.035 | 264.585 | 85.035 | 16.655 | 1.503 |
| | | | | *L-* | *S-* | -60.232 | 78.333 | 74.769 | 4.085 | 0.510 |
| | 109 | 90.000 | 45.000 | *R/L-* | *S-* | ±9.216 | 167.077 | 83.497 | 9.944 | 0.988 |
| | | | | | *L-* | ∓56.247 | 240.904 | 53.991 | 14.512 | 3.156 |
| | | | | | | **∓90.000** | **250.529** | **45.000** | **2∞** | **3.239** |
| 90.000 | 71 | 90.000 | 0.000 | *R/L-* | *S-* | ±90.000 | 70.529 | 90.000 | 4.264 | 0.461 |
| | | | | | *L-* | **∓90.000** | **289.471** | **90.000** | **2∞** | **3.197** |
| | | | | | | ∓39.734 | 264.227 | 90.000 | 16.341 | 1.497 |
| | 109 | 0.000 | 90.000 | *R/L-* | *S-* | ±90.000 | 109.471 | 90.000 | 3.309 | 1.368 |
| | | | | | *L-* | **∓90.000** | **250.529** | **90.000** | **2∞** | **2.290** |
| | 180 | 35.264 | 54.736 | *R/L-* | *M-* | **-90.000** | **180.000** | **90.000** | **∞** | **1.829** |
| | | | | | | **90.000** | **180.000** | **90.000** | **∞** | **1.829** |

Here and hereinafter the next abridgements are assumed: Int.angl. – interfacial angle.

Table B.2. Energy density and structure of the plane Bloch DW in (110)-plate

| Int.angl. | DW type | | | M turn type | | DW energy density and structure | | | | |
|---|---|---|---|---|---|---|---|---|---|---|
| $\phi$, deg. | $2\alpha$, deg. | $\beta$, deg. | $\gamma$, deg. | Rotation | Path | $\lambda$, deg. | $\Delta\tilde{\varphi}$, deg. | $\psi$, deg. | $\bar{\delta}$, arb.units. | $\bar{\sigma}/\sin\psi$, arb.units. |
| 0.000 | 71 | 90.000 | 90.000 | *R/L-* | *S-* | $\pm$*35.217* | *101.603* | *54.783* | *3.650* | *0.738* |
| | | | | | *L-* | $\mp$*35.616* | *258.947* | *54.385* | *13.028* | *1.899* |
| | 109 | 90.000 | 90.000 | *R/L-* | *S-* | $\pm$*8.481* | *168.093* | *81.519* | *10.143* | *0.994* |
| | 180 | 90.000 | 90.000 | *R/L-* | *M-* | *7.212* | *180.000* | *82.788* | *8.135* | *2.004* |
| | | | | | | *-7.212* | *180.000* | *82.788* | *8.135* | *2.004* |
| 60.000 | 71 | 60.000 | 45.000 | *R-* | *L-* | -39.890 | 264.414 | 86.389 | 16.502 | 1.500 |
| | | | | | | -89.046 | 289.464 | 60.554 | 24.056 | 3.683 |
| | | | | *L-* | *S-* | -67.577 | 74.828 | 74.248 | 4.163 | 0.497 |
| | | | | | *L-* | 39.380 | 263.801 | 45.275 | 15.987 | 2.107 |
| | 109 | 45.000 | 60.000 | *R-* | *S-* | 3.649 | 174.847 | 42.482 | 12.083 | 1.509 |
| | | | | | *L-* | -63.489 | 244.648 | 82.427 | 12.026 | 2.508 |
| | | | | *L-* | *S-* | -20.809 | 151.798 | 61.095 | 7.614 | 1.183 |
| 90.000 | 71 | 0.000 | 90.000 | *R/L-* | *S-* | $\pm$90.000 | 70.529 | 90.000 | 4.264 | 0.461 |
| | | | | | *L-* | **$\mp$90.000** | **289.471** | **90.000** | **2∞** | **3.197** |
| | | | | | | $\mp$42.813 | 267.728 | 42.813 | 20.792 | 2.263 |
| | 109 | 90.000 | 0.000 | *R/L-* | *S-* | $\pm$10.146 | 165.799 | 90.000 | 9.710 | 0.981 |
| | | | | | *L-* | **$\mp$90.000** | **250.529** | **90.000** | **2∞** | **2.290** |
| | 180 | 54.736 | 35.264 | *R/L-* | *M-* | **90.000** | **180.000** | **90.000** | **∞** | **1.829** |
| | | | | | | **-90.000** | **180.000** | **90.000** | **∞** | **1.829** |
| | | | | | | 33.555 | 180.000 | 61.240 | 12.301 | 2.103 |
| | | | | | | -33.555 | 180.000 | 61.240 | 12.301 | 2.103 |

Table B.3. Energy density and structure of the plane Bloch DW in (111)-plates

| Int.angl. | DW type | | | **M** turn type | | DW energy density and structure | | | | |
|---|---|---|---|---|---|---|---|---|---|---|
| $\phi$, deg. | $2\alpha$, deg. | $\beta$, deg. | $\gamma$, deg. | Rotation | Path | $\lambda$, deg. | $\Delta\tilde{\varphi}$, deg. | $\psi$, deg. | $\bar{\delta}$, arb.units. | $\bar{\sigma}/\sin\psi$, arb.units. |
| 35.264 | 71 | 90.000 | 54.736 | *R/L-* | *S-* | $\pm$45.000 | 90.000 | 54.736 | 3.848 | 0.667 |
| | | | | | *L-* | $\mp$37.881 | 261.939 | 59.911 | 14.663 | 1.743 |
| | | | | | | ***$\mp$90.000*** | ***289.471*** | ***35.264*** | ***2∞*** | ***5.538*** |
| | 109 | 54.736 | 90.000 | *R-* | *S-* | *3.236* | *175.428* | *51.500* | *12.357* | *1.309* |
| | | | | | *L-* | -55.894 | 240.697 | 69.371 | 15.067 | 2.734 |
| | | | | *L-* | *S-* | -16.188 | 157.696 | 70.924 | 8.453 | 1.060 |
| 90.000 | 71 | 35.264 | 54.736 | *R/L-* | *S-* | *$\pm$90.000* | *70.529* | *90.000* | *4.264* | *0.461* |
| | | | | | *L-* | ***$\mp$90.000*** | ***289.471*** | ***90.000*** | ***2∞*** | ***3.197*** |
| | | | | | | $\mp$41.383 | 266.147 | 52.221 | 18.299 | 1.907 |
| | 109 | 54.736 | 35.264 | *R/L-* | *S-* | $\pm$11.389 | 164.102 | 55.530 | 9.421 | 1.192 |
| | | | | | | $\pm$90.000 | 109.471 | 90.000 | 3.309 | 1.368 |
| | | | | | *L-* | **$\mp$90.000** | **250.529** | **90.000** | **2∞** | **2.290** |
| | 180 | 90.000 | 0.000 | *R/L-* | *M-* | ***30.000*** | ***180.000*** | ***90.000*** | ***∞*** | ***1.829*** |
| | | | | | | ***-30.000*** | ***180.000*** | ***90.000*** | ***∞*** | ***1.829*** |
| | | | | | | ***90.000*** | ***180.000*** | ***90.000*** | ***∞*** | ***1.829*** |
| | | | | | | ***-90.000*** | ***180.000*** | ***90.000*** | ***∞*** | ***1.829*** |
| | 180 | 19.471 | 70.529 | *R/L-* | *M-* | ***-90.000*** | ***180.000*** | ***90.000*** | ***∞*** | ***1.829*** |
| | | | | | | ***90.000*** | ***180.000*** | ***90.000*** | ***∞*** | ***1.829*** |

Table B.4. Energy density and structure of the plane Bloch DW in (112)-plate

| Int.angl. | DW type | | | **M** turn type | | DW energy density and structure | | | | |
|---|---|---|---|---|---|---|---|---|---|---|
| $\phi$, deg. | $2\alpha$, deg. | $\beta$, deg. | $\gamma$, deg. | Rota-tion | Path | $\lambda$, deg. | $\Delta\tilde{\varphi}$, deg. | $\psi$, deg. | $\bar{\delta}$, arb.units. | $\bar{\sigma}/\sin\psi$, arb.units. |
| 30.000 | 71 | 73.221 | 65.905 | *R-* | *S-* | 31.406 | 107.224 | 45.760 | 3.568 | 0.885 |
| | | | | | *L-* | -38.543 | 262.773 | 71.709 | 15.218 | 1.581 |
| | | | | *L-* | *S-* | -47.953 | 87.195 | 63.271 | 3.901 | 0.595 |
| | 109 | 65.905 | 73.221 | *R-* | *S-* | 5.180 | 172.695 | 61.003 | 11.285 | 1.146 |
| | | | | | *L-* | -54.729 | 239.996 | 61.881 | 27.308 | 2.926 |
| | | | | *L-* | *S-* | -12.994 | 161.932 | 78.284 | 9.077 | 1.008 |
| 54.736 | 71 | 90.000 | 35.264 | *R/L-* | *S-* | ±75.348 | 72.324 | 56.043 | 4.221 | 0.564 |
| | | | | | *L-* | ∓39.020 | 263.363 | 68.685 | 15.646 | 1.608 |
| | | | | | | **∓90.000** | **289.471** | **54.736** | **2∞** | **3.916** |
| | 109 | 35.264 | 90.000 | *R-* | *M-* | **0.000** | **180.000** | **35.264** | **2∞** | **1.886** |
| | | | | | *L-* | -62.402 | 244.147 | 82.334 | 12.067 | 2.514 |
| | | | | *L-* | *S-* | -30.159 | 140.885 | 65.423 | 5.882 | 1.230 |
| | 180 | 48.190 | 61.875 | *R/L-* | *M-* | -33.877 | 180.000 | 76.604 | 12.090 | 1.897 |
| 73.221 | 71 | 30.000 | 65.905 | *R-* | *L-* | -41.193 | 265.931 | 62.512 | 18.038 | 1.697 |
| | | | | | | -89.304 | 289.467 | 73.851 | 25.611 | 3.335 |
| | | | | *L-* | *S-* | -78.089 | 71.709 | 84.047 | 4.236 | 0.468 |
| | | | | | *L-* | 42.657 | 267.559 | 33.643 | 20.456 | 2.768 |
| | 109 | 65.905 | 30.000 | *R-* | *S-* | 8.733 | 167.745 | 63.427 | 10.074 | 1.099 |
| | | | | | *L-* | -89.351 | 250.525 | 73.499 | 28.622 | 2.391 |
| | | | | *L-* | *S-* | -12.062 | 163.190 | 70.189 | 9.273 | 1.046 |

Table B.4 (continue).

| Int.angl. | DW type | | | **M** turn type | | DW energy density and structure | | | | |
|---|---|---|---|---|---|---|---|---|---|---|
| $\phi$, deg. | $2\alpha$, deg. | $\beta$, deg. | $\gamma$, deg. | Rotation | Path | $\lambda$, deg. | $\Delta\tilde{\varphi}$, deg. | $\psi$, deg. | $\bar{\delta}$, arb.units. | $\bar{\sigma}/\sin\psi$, arb.units. |
| 90.000 | 71 | 54.736 | 35.264 | *R/L-* | *S-* | ±90.000 | 70.529 | 90.000 | 4.264 | 0.461 |
| | | | | | *L-* | **∓90.000** | **289.471** | **90.000** | **2∞** | **3.197** |
| | | | | | | ∓40.417 | 265.036 | 63.924 | 17.079 | 1.668 |
| | 109 | 35.264 | 54.736 | *R/L-* | *S-* | ±90.000 | 109.471 | 90.000 | 3.309 | 1.368 |
| | | | | | *L-* | **∓90.000** | **250.529** | **90.000** | **2∞** | **2.290** |
| | 180 | 70.529 | 19.471 | *R/L-* | *M-* | **90.000** | **180.000** | **90.000** | ∞ | **1.829** |
| | | | | | | **-90.000** | **180.000** | **90.000** | ∞ | **1.829** |
| | | | | | | 30.469 | 180.000 | 73.304 | 17.257 | 1.910 |
| | | | | | | -30.469 | 180.000 | 73.304 | 17.257 | 1.910 |
| | 180 | 0.000 | 90.000 | *R/L-* | *M-* | ***90.000*** | ***180.000*** | ***90.000*** | ∞ | ***1.829*** |
| | | | | | | ***-90.000*** | ***180.000*** | ***90.000*** | ∞ | ***1.829*** |

Table B.5. Energy density and structure of the plane Bloch DW in (210)-plate

| Int.angl. | DW type | | | **M** turn type | | DW energy density and structure | | | | |
|---|---|---|---|---|---|---|---|---|---|---|
| $\phi$, deg. | $2\alpha$, deg. | $\beta$, deg. | $\gamma$, deg. | Rota-tion | Path | $\lambda$, deg. | $\Delta\tilde{\varphi}$, deg. | $\psi$, deg. | $\bar{\delta}$, arb.units. | $\bar{\sigma}/\sin\psi$, arb.units. |
| 18.435 | 71 | 71.565 | 90.000 | *R-* | *S-* | 25.437 | 117.449 | 46.129 | 3.441 | 0.964 |
| | | | | | *L-* | -38.250 | 262.406 | 70.185 | 14.967 | 1.599 |
| | | | | *L-* | *S-* | -45.000 | 90.000 | 63.435 | 3.848 | 0.609 |
| | 109 | 90.000 | 71.565 | *R/L-* | *S-* | ±8.616 | 167.906 | 81.829 | 10.106 | 0.993 |
| 50.769 | 71 | 50.769 | 63.435 | *R-* | *L-* | -87.347 | 289.413 | 52.951 | 18.988 | 4.078 |
| | | | | | | -39.978 | 264.518 | 85.509 | 16.595 | 1.501 |
| | | | | *L-* | *S-* | -62.439 | 77.153 | 74.452 | 4.111 | 0.506 |
| | | | | | *L-* | 37.616 | 261.601 | 27.500 | 14.452 | 3.272 |
| | 109 | 63.435 | 50.769 | *R-* | *S-* | 6.138 | 171.353 | 59.185 | 10.896 | 1.158 |
| | | | | | *L-* | -58.607 | 242.230 | 72.126 | 12.801 | 2.650 |
| | | | | *L-* | *S-* | -13.481 | 161.279 | 73.294 | 8.977 | 1.032 |
| | 180 | 90.000 | 39.232 | *R/L-* | *M-* | -27.172 | 180.000 | 73.213 | 12.859 | 1.921 |
| | | | | | | 27.172 | 180.000 | 73.213 | 12.859 | 1.921 |
| | | | | | | **-90.000** | **180.000** | **50.769** | **∞** | **2.362** |
| | | | | | | **90.000** | **180.000** | **50.769** | **∞** | **2.362** |
| | 180 | 43.089 | 75.037 | *R/L-* | *M-* | -35.034 | 180.000 | 76.414 | 11.455 | 1.907 |

Table B.5 (continue)

| Int.angl. | DW type | | | **M** turn type | | DW energy density and structure | | | | |
|---|---|---|---|---|---|---|---|---|---|---|
| $\phi$, deg. | $2\alpha$, deg. | $\beta$, deg. | $\gamma$, deg. | Rota-tion | Path | $\lambda$, deg. | $\Delta\tilde{\varphi}$, deg. | $\psi$, deg. | $\bar{\delta}$, arb.units. | $\bar{\sigma}/\sin\psi$, arb.units. |
| 71.565 | 71 | 18.435 | 90.000 | *R-* | *L-* | -89.100 | 289.465 | 72.465 | 24.346 | 3.363 |
| | | | | | | -41.519 | 266.301 | 59.954 | 18.494 | 1.744 |
| | | | | *L-* | *S-* | -77.351 | 71.861 | 84.214 | 4.232 | 0.469 |
| | | | | | *L-* | 44.803 | 269.802 | 26.368 | 32.394 | 3.651 |
| | 109 | 90.000 | 18.435 | *R/L-* | *S-* | ±9.940 | 166.082 | 86.871 | 9.761 | 0.983 |
| | | | | | *L-* | **∓90.000** | **250.529** | **71.565** | **2∞** | **2.414** |
| | 71 | 71.565 | 26.565 | *R-* | *L-* | -89.771 | 289.471 | 71.642 | 31.056 | 3.370 |
| | | | | | | -39.798 | 264.304 | 87.676 | 16.406 | 1.498 |
| | | | | *L-* | *S-* | -77.876 | 71.752 | 75.951 | 4.235 | 0.480 |
| | | | | | *L-* | 39.651 | 264.128 | 63.560 | 16.256 | 1.672 |
| | 109 | 26.565 | 71.565 | *R-* | *M-* | **0.000** | **180.000** | **26.564** | **2∞** | **2.434** |
| | | | | | *L-* | -87.404 | 250.473 | 74.014 | 21.857 | 2.402 |
| | | | | *L-* | *S-* | -59.607 | 117.239 | 79.645 | 3.687 | 1.333 |

Table B.6. Energy density and structure of the plane Bloch DW in (123)-plate

| Int.angl. | DW type | | | **M** turn type | | DW energy density and structure | | | | |
|---|---|---|---|---|---|---|---|---|---|---|
| $\phi$, deg. | $2\alpha$, deg. | $\beta$, deg. | $\gamma$, deg. | Rota-tion | Path | $\lambda$, deg. | $\Delta\tilde{\varphi}$, deg. | $\psi$, deg. | $\bar{\delta}$, arb.units. | $\bar{\sigma}/\sin\psi$, arb.units. |
| 19.107 | 71 | 79.107 | 74.499 | *R-* | *S-* | 31.251 | 107.467 | 49.325 | 3.565 | 0.838 |
| | | | | | *L-* | -37.757 | 261.781 | 64.585 | 14.563 | 1.671 |
| | | | | *L-* | *S-* | -42.495 | 92.617 | 60.068 | 3.800 | 0.643 |
| | 109 | 74.499 | 79.107 | *R-* | *S-* | 6.264 | 171.176 | 68.360 | 10.849 | 1.069 |
| | | | | *L-* | *S-* | -11.156 | 164.419 | 85.447 | 9.473 | 0.985 |
| 40.893 | 71 | 67.792 | 57.689 | *R-* | *S-* | 36.902 | 99.327 | 40.875 | 3.686 | 0.903 |
| | | | | | *L-* | -39.144 | 263.514 | 79.393 | 15.762 | 1.524 |
| | | | | | | -86.963 | 289.395 | 42.706 | 18.316 | 4.820 |
| | | | | *L-* | *S-* | -54.360 | 82.051 | 66.790 | 4.005 | 0.553 |
| | | | | | *L-* | 36.189 | 259.725 | 41.290 | 13.409 | 2.323 |
| | 109 | 57.689 | 67.792 | *R-* | *S-* | 4.275 | 173.966 | 53.887 | 11.722 | 1.252 |
| | | | | | *L-* | -56.661 | 241.144 | 70.260 | 14.028 | 2.705 |
| | | | | *L-* | *S-* | -15.193 | 159.003 | 71.474 | 8.641 | 1.051 |
| | 180 | 77.396 | 51.887 | *R/L-* | *M-* | -28.291 | 180.000 | 80.438 | 14.090 | 1.859 |
| | 180 | 49.107 | 90.000 | *R/L-* | *M-* | -32.189 | 180.000 | 81.296 | 13.484 | 1.857 |
| 55.462 | 71 | 79.107 | 36.699 | *R-* | *L-* | -39.273 | 263.671 | 77.726 | 15.885 | 1.532 |
| | | | | | | -89.635 | 289.470 | 55.547 | 28.784 | 3.880 |
| | | | | *L-* | *S-* | -64.873 | 75.981 | 64.339 | 4.137 | 0.535 |
| | | | | | *L-* | 38.899 | 263.213 | 59.796 | 15.535 | 1.735 |
| | 109 | 36.699 | 79.107 | *R-* | *S-* | 0.444 | 179.373 | 36.279 | 17.020 | 1.821 |
| | | | | | *L-* | -62.534 | 244.209 | 82.343 | 12.059 | 2.513 |
| | | | | *L-* | *S-* | -28.697 | 142.492 | 64.464 | 6.140 | 1.225 |

Table B.6 (continue).

| Int.angl. | DW type | | | **M** turn type | | DW energy density and structure | | | | |
|---|---|---|---|---|---|---|---|---|---|---|
| $\phi$, deg. | $2\alpha$, deg. | $\beta$, deg. | $\gamma$, deg. | Rota-tion | Path | $\lambda$, deg. | $\Delta\tilde{\varphi}$, deg. | $\psi$, deg. | $\bar{\delta}$, arb.units. | $\bar{\sigma}/\sin\psi$, arb.units. |
| 67.792 | 71 | 40.893 | 57.689 | *R-* | *L-* | -40.671 | 265.331 | 70.912 | 17.377 | 1.587 |
| | | | | | | -89.077 | 289.464 | 68.547 | 24.222 | 3.445 |
| | | | | *L-* | *S-* | -73.933 | 72.695 | 81.142 | 4.213 | 0.475 |
| | | | | | *L-* | 41.389 | 266.154 | 35.212 | 18.307 | 2.615 |
| | 109 | 57.689 | 40.893 | *R-* | *S-* | 7.338 | 169.678 | 54.660 | 10.483 | 1.211 |
| | | | | | *L-* | -88.261 | 250.504 | 68.803 | 23.805 | 2.467 |
| | | | | | | -66.105 | 245.765 | 82.585 | 12.108 | 2.496 |
| | | | | *L-* | *S-* | -14.127 | 160.417 | 64.779 | 8.848 | 1.095 |
| | 180 | 90.000 | 22.208 | *R/L-* | *M-* | -29.405 | 180.000 | 79.306 | 16.671 | 1.862 |
| | | | | | | 29.405 | 180.000 | 79.306 | 16.671 | 1.862 |
| | | | | | | **-90.000** | **180.000** | **67.792** | **∞** | **1.976** |
| | | | | | | **90.000** | **180.000** | **67.792** | **∞** | **1.976** |
| | 180 | 29.206 | 72.025 | *R/L-* | *M-* | -82.740 | 180.000 | 74.655 | 10.573 | 1.941 |
| 79.107 | 71 | 55.462 | 36.699 | *R-* | *L-* | -40.325 | 264.928 | 71.945 | 16.975 | 1.576 |
| | | | | | | -89.774 | 289.471 | 79.237 | 31.121 | 3.255 |
| | | | | *L-* | *S-* | -81.899 | 71.071 | 83.846 | 4.251 | 0.466 |
| | | | | | *L-* | 40.336 | 264.940 | 56.325 | 16.987 | 1.800 |
| | 109 | 36.699 | 55.462 | *R-* | *S-* | 8.113 | 168.603 | 34.872 | 10.248 | 1.722 |
| | | | | | *L-* | -89.184 | 250.523 | 79.773 | 27.500 | 2.330 |
| | | | | *L-* | *S-* | -65.800 | 114.359 | 81.008 | 3.535 | 1.350 |

Table B.6 (continue).

| Int.angl. | DW type | | | **M** turn type | | DW energy density and structure | | | | |
|---|---|---|---|---|---|---|---|---|---|---|
| $\phi$, deg. | $2\alpha$, deg. | $\beta$, deg. | $\gamma$, deg. | Rota-tion | Path | $\lambda$, deg. | $\Delta\tilde{\varphi}$, deg. | $\psi$, deg. | $\bar{\delta}$, arb.units. | $\bar{\sigma}/\sin\psi$, arb.units. |
| 79.107 | 71 | 19.107 | 74.499 | *R*- | *L*- | -41.770 | 266.583 | 54.630 | 18.869 | 1.856 |
| | | | | | | -89.582 | 289.470 | 79.509 | 28.113 | 3.254 |
| | | | | *L*- | *S*- | -82.381 | 71.009 | 86.444 | 4.253 | 0.464 |
| | | | | | *L*- | 43.240 | 268.185 | 35.133 | 21.847 | 2.696 |
| | 109 | 74.499 | 19.107 | *R*- | *S*- | 9.603 | 166.545 | 72.840 | 9.845 | 1.027 |
| | | | | | *L*- | -89.812 | 250.528 | 79.158 | 34.695 | 2.332 |
| | | | | *L*- | *S*- | -10.863 | 164.818 | 76.888 | 9.540 | 1.008 |

**Reference**

[1] R. Vakhitov, A Yumaguzin, J. Magn. Magn. Mater. 215-216, 52 (2000)

[2] R.M. Sabitov, R.M. Vakhitov, Sov. Phys. Izv. Vuzov. Fizika, 8 (1988) 51

[3] B.A. Lilley. Phil.Mag. **41** (1950) 792.

[4] L.D. Landau, E.M. Lifshitz. Sow.Phys. **8** (1935) 153.

[5] A. Hubert. Theory of Domain Walls. Springer, New York, 1974.

[6] Hubert A., Shafer R. Magnetic domains. The analysis of magnetic microstructures. – Berlin, Springer-Verlag, 1998.

[7] A. Aharoni, J.P. Jakubovics. J.Magn.Magn.Mater. **104-107** (1992) 353

[8] F.B. Humphrey, M. Redjdal. J.Magn.Magn.Mater. **133** (1994) 11

[9] V.L. Vlasko-Vlasov, L.M. Dedukh, V.I. Nikitenko. JETP **71** (1976) 2291

[10] J. Peters, W. Treimer. J.Magn.Magn.Mater. **241** (2002) 240

[11]. O. Antonyuk, V.F Kovalenko, O.V. Tychko. Journ.Alloys and Compaund. **369** (2004) 112

[12] V.A. Gurevich et al., sov. phys.: sol. stat. phys. **19** (1977) 761

[13] O. Scharpf, R. Seifert, H. Strothmann. J.Magn.Magn.Mater. **13** (1979) 239

[14] Stupakiewicz A., Maziewski A., Davidenko I., Zablotskii V. Phys.Rev.B. **64** (2001) 064405

[15] A. Szewczyk, Z. Henkie. J.Magn.Magn.Mater. **81** (1989) 277

[16] L.A. Pamyatnykh, G.S. Kandaurova, M.A. Shamsutdinov, V.V. Plavski, B.N. Filippov. J.Magn.Magn.Mater. **234** (2001) 469

[17] L.M. Dedukh, V.I. Nikitenko, V.T. Synogach. JETP Letters. **45** (1987) 491

[18] V. V. Randoshkin, A. M. Saletskii, N. N. Usmanov, D. B. Chopornyak. Physics of the Solid State **44** (2002) 899

[19] V. Baryakhtar, V. Lvov, D. Yablonsky, JETP 87 (1984) 1863

[20] B. M. Tanygin, O. V. Tychko, Physica B: Condensed Matter, 404, 21, 4018-4022 (2009).

[21] J.F. Dillon, J.P. Remeika, E.M. Gyorgy. Appl. Phys. Lett. **9** (1966) 147

[22] S.T. Chui and V.N. Ryzhov J. Magn. Magn. Mater. **182** (1998) 25

[23] S.A. D’yachenko et al., Physics of Solid State, 2008, 50, 1, 32-42.